\documentclass{aa}

\usepackage{natbib}
\usepackage{hyperref}
\hypersetup{colorlinks=true,
            linkcolor=blue,
            citecolor=blue,
            filecolor=blue,
            urlcolor=blue,
            }

\usepackage{graphicx}	
\usepackage{amsmath}	
\usepackage{amssymb}	
\usepackage{multicol}        
\usepackage{bm}		
\usepackage{xcolor}
\usepackage{subcaption}
\usepackage[labelfont=bf]{caption}

\newcommand{\gx}{\textsc{Gadget-X}}

\newcommand{\Mcrit}{{\ifmmode{M_{\rm 200c}}\else{$M_{\rm 200c}$}\fi}}
\newcommand{\Rcrit}{{\ifmmode{R_{\rm 200c}}\else{$R_{\rm 200c}$}\fi}}
\newcommand{\Rhost}{{\ifmmode{R_{\rm host}}\else{$R_{\rm host}$}\fi}}
\newcommand{\Mmean}{{\ifmmode{M_{\rm 200m}}\else{$M_{\rm 200m}$}\fi}}

\newcommand{\hGpc}{{\ifmmode{h^{-1}{\rm Gpc}}\else{$h^{-1}$Gpc}\fi}}
\newcommand{\hMpc}{{\ifmmode{h^{-1}{\rm Mpc}}\else{$h^{-1}$Mpc}\fi}}
\newcommand{\hkpc}{{\ifmmode{h^{-1}{\rm kpc}}\else{$h^{-1}$kpc}\fi}}
\newcommand{\hMsun}{{\ifmmode{h^{-1}{\rm {M_{\odot}}}}\else{$h^{-1}{\rm{M_{\odot}}}$}\fi}}
\newcommand{\Mstar}{{\ifmmode{M_{*}}\else{$M_{*}$}\fi}}
\newcommand{\Mhalo}{{\ifmmode{M_{\rm Halo}}\else{$M_{\rm Halo}$}\fi}}
\newcommand{\Ngal}{{\ifmmode{N_{\rm gal}}\else{$N_{\rm gal}$}\fi}}
\newcommand{\Msun}{{\ifmmode{{\rm {M_{\odot}}}}\else{${\rm{M_{\odot}}}$}\fi}}
\newcommand{\ltsima}{$\; \buildrel < \over \sim \;$}
\newcommand{\gtsima}{$\; \buildrel > \over \sim \;$}
\newcommand{\lsim}{\lower.5ex\hbox{\ltsima}}
\newcommand{\gsim}{\lower.5ex\hbox{\gtsima}}

\newcommand{\Sec}[1]{Section~\ref{#1}}

\newcommand{\Fig}[1]{Fig.~\ref{#1}}

\defcitealias{Contreras-Santos2024}{Paper I}

\usepackage[T1]{fontenc}
\usepackage{ae,aecompl}

\usepackage{newtxtext,newtxmath}
\usepackage{natbib}
\bibpunct{(}{)}{;}{a}{}{,} 

\begin{document} 

   \titlerunning{Origin of the ICL}
   \title{The origin of the intra-cluster light in The Three Hundred simulations}

   \author{A. Contreras-Santos
          \inst{1}
          \and
          A. Knebe \inst{1,2,3}
          \and
          W. Cui \inst{1,2,4}
          \and
          I. Alonso Asensio \inst{5,6}
          \and
          C. Dalla Vecchia \inst{5,6}
          \and
          R. Haggar \inst{7,8}
          \and
          R. A. Mostoghiu Paun \inst{9,10}
          \and
          F. R. Pearce \inst{11}
          \and
          E. Rasia \inst{12,13,14}
          \and
          G. Martin \inst{11}
          \and
          S. E. Nuza \inst{15}
          \and
          G. Yepes \inst{1,2}
          }

   \institute{Departamento de F\'isica Te\'{o}rica, M\'{o}dulo 15, Facultad de Ciencias, Universidad Aut\'{o}noma de Madrid, 28049 Madrid, Spain
         \and
             Centro de Investigaci\'{o}n Avanzada en F\'isica Fundamental (CIAFF), Facultad de Ciencias, Universidad Aut\'{o}noma de Madrid, 28049 Madrid, Spain
         \and
             International Centre for Radio Astronomy Research, University of Western Australia, 35 Stirling Highway, Crawley, Western Australia 6009, Australia
         \and
             Institute for Astronomy, University of Edinburgh, Royal Observatory, Blackford Hill, Edinburgh EH9 3HJ, UK
         \and
             Instituto de Astrof\'{i}sica de Canarias, C/V\'{i}a L\'{a}ctea s/n, 38205 La Laguna, Tenerife, Spain
         \and
             Departamento de Astrof\'{i}sica, Universidad de La Laguna, Av. Astrof\'{i}sico Francisco Sanchez s/n, E-38206 La Laguna, Tenerife, Spain
         \and
             Department of Physics and Astronomy, University of Waterloo, Waterloo, Ontario N2L 3G1, Canada
        \and
             Waterloo Centre for Astrophysics, University of Waterloo, Waterloo, Ontario N2L 3G1, Canada
        \and
            Centre for Astrophysics \& Supercomputing, Swinburne University of Technology, 1 Alfred St, Hawthorn, VIC 3122, Australia
        \and
            ARC Centre of Excellence for Dark Matter Particle Physics (CDM), Australia
        \and
            School of Physics \& Astronomy, University of Nottingham, Nottingham NG7 2RD, United Kingdom
        \and
            INAF – Osservatorio Astronomico di Trieste, via Tiepolo 11, I34131 Trieste, Italy
        \and
            IFPU – Institute for Fundamental Physics of the Universe, via Beirut 2, 34151, Trieste, Italy
        \and
            Department of Physics; University of Michigan, 450 Church St, Ann Arbor, MI 48109, USA
        \and
            Instituto de Astronom\'{\i}a y F\'{\i}sica del Espacio (IAFE, CONICET-UBA), CC 67, Suc. 28, 1428 Buenos Aires, Argentina
             }

\date{Received January 1, 2020; accepted January 1, 2021}

 
  \abstract{
  We investigate the origin and formation mechanisms of the intra-cluster light (ICL) in \textsc{The Three Hundred} simulations, a set of 324 hydrodynamically resimulated massive galaxy clusters. The ICL, a diffuse component comprised of stars not bound to any individual galaxy, serves as a critical tracer of cluster formation and evolution. Using two implementations of the hydrodynamics, \textsc{Gadget-X} and \textsc{Gizmo-Simba}, we identify the stellar particles that constitute the ICL at $z=0$ and trace them back in time to the moments when they were formed and accreted into the ICL. 
  Our analysis reveals that, across our 324 clusters, half of the present-day ICL mass is typically in place between $z \sim 0.2$ and 0.5. The main ICL formation channel is the stripping of stars from subhalos after their infall into the host cluster. Within this channel, 65-80 per cent of the ICL comes from objects with stellar (infall) masses above $10^{11}$ M$_\odot$, corresponding to massive galaxies, groups and clusters. Considering also the ratio of the infalling halo to the total cluster mass, we see that a median of 35 per cent of the mass is brought in major merger events, although the percentage varies significantly across clusters (15-55 per cent). Additional contributions come from minor mergers (25-35 per cent) and smooth accretion (20-50 per cent). The infall redshift of the primary contributors is generally below $z\leq1$, with smaller fractions arriving at redshifts between 1 and 2. 
  Regarding other formation channels, we find minor contributions from stars formed in subhalos after their infall and stars stripped while their contributing halo remains outside the host cluster (and can eventually fall inside or stay outside). Finally, for our two sets of simulations, we find medians of 12 (\textsc{Gadget-X}) and 2 (\textsc{Gizmo-Simba}) per cent of the ICL mass formed in-situ, that is, directly as part of the diffuse component. However, this component can be attributed to stripping of gas in high-velocity infalling satellite galaxies.  
  }


   \maketitle
%

\section{Introduction}

The intra-cluster light (ICL) is a diffuse component in galaxy clusters, coming from stars that are not bound to any individual galaxy but to the potential of the cluster itself. It was first theorised and then discovered more than 70 years ago by \citet{Zwicky1937,Zwicky1951}. However, over the past two decades, interest in the ICL has grown significantly due to the advent of modern instruments that allow us to observe this faint light from galaxy clusters. The formation of the ICL is closely linked to that of the cluster, and so it carries information about the past, present and future history of the cluster \citep[see, e.g., reviews by][]{Contini2021,Montes2019,Montes2022}. Thus, understanding the ICL becomes essential for a complete understanding of galaxy clusters and their role in galaxy evolution.

Despite its importance, defining and quantifying the ICL remains a major challenge, both observationally and in simulations. Theoretically, the ICL consists of stars unbound from galaxies but bound to the cluster. However, in practice, separating the ICL from the brightest cluster galaxy (BCG), usually located in the centre, is not straightforward \citep[see reviews by][]{Montes2019,Montes2022,Contini2021}. Most observational studies often rely on surface brightness thresholds \citep[e.g.][]{Feldmeier2004,MontesTrujillo2014,Burke2015,Furnell2021,Montes2021} or multi-component profile fitting to isolate the ICL \citep[e.g.][]{Gonzalez2005,Janowiecki2010,Spavone2018,Montes2021,Joo2023,Ragusa2023}, while others simply consider the joint component BCG+ICL \citep[e.g.][]{Gonzalez2013,DeMaio2018,Zhang2019,DeMaio2020}. Meanwhile, simulations allow for more theoretical approaches based on gravitational binding energy, stellar kinematics or phase space structure \citep[e.g.][]{Dolag2010,Cui2014,Remus2017,AlonsoAsensio2020,Montenegro-Taborda2023}, although the task of separating the different gravitational potentials of galaxies and host clusters is also not trivial \citep{Canas2020}. These methodological differences contribute to significant variation in reported ICL fractions. 
In addition to methodological differences, numerical effects can also influence ICL estimates in simulations. Recent work by \citet{Martin2024} shows that poorly resolved dark matter halos might be over-stripped in cosmological simulations, which can be particularly important for low-mass systems. 
Comparative efforts, such as \citet{Brough2024}, who examine BCG and ICL fractions across multiple methods using both observed and simulated data, are crucial for understanding the implications of different approaches. Promisingly, recent studies have also explored the use of machine learning algorithms to achieve a more efficient and unbiased separation \citep{Marini2022,Marini2024,Canepa2025}.

At $z=0$, both observational and theoretical studies broadly agree that the ICL represents a non-negligible fraction of clusters' stellar mass, typically ranging between 10 and 40 per cent, though with significant scatter depending on definitions and methods \citep[see, e.g.][]{Montes2022,Brough2024}. Going to higher redshifts, the build-up of ICL over cosmic time remains poorly constrained. From the point of view of observations, there is a trend of increasing ICL fraction with decreasing redshift. With a collection of previous measurements ranging between $z\sim0$ and $z\sim1.2$, \citet{Montes2022} showed that the ICL fraction remains roughly constant at $z>0.6$, with values below 10 per cent. At $z\sim0.6$ the ICL clearly starts to build-up, but the steepness of this increase depends on the method used to define it, with surface brightness cuts showing the strongest dependency. In opposition to this, \citet{Joo2023} conclude that the ICL is already abundant at $z>1$, finding a mean ICL fraction of $\sim17$ per cent for ten clusters at $1<z<2$. Additionally, \citet{Werner2023} report diffuse light detections in two proto-clusters at $z \sim 2$, implying that the ICL component may begin forming earlier than previously expected.

On the theoretical side, predictions for the evolution of the ICL fraction with redshift also vary significantly. Simulation-based studies such as those by \citet{Rudick2011} and \citet{Tang2018} suggest that the amount of ICL is negligible ($< 5\%$) at $z>1$, and then increases notably up to 15-30 per cent at $z=0$. A recent semi-empirical model by \citet{Fu2024} supports a similar evolutionary trend. In contrast, semi-analytic models by \citet{Contini2023b} predict a nearly constant ICL fraction over time, with no clear dependence on redshift. This flat evolution is more in line with the observational results of \citet{Joo2023}, although their modelled ICL fractions seem to be slightly higher than the observations. Overall, the timing and rate of ICL growth remain uncertain, and further research is needed to clarify the picture.

The significance of the ICL fraction at high redshifts is closely linked to the formation of this component. A weak or nonexistent correlation of the fraction with redshift would indicate that most of the ICL formation happened at high redshifts. 
Another method used in observations to address the issue of ICL formation is through the colour, metallicity and age of its stellar population \citep[e.g.][]{Mihos2017,Iodice2017,Morishita2017,DeMaio2018,MontesTrujillo2018,Edwards2019,Kluge2024}. This can provide information not only about the timescales of ICL formation but also about the dominant formation channels \citep[see][for a review]{Montes2019}. For instance, the absence or presence of gradients in the ICL colour profile can give hints about the relevance of gradual stripping against major mergers in ICL formation. Apart from stripping and mergers, other proposed mechanisms to produce ICL include the `pre-processing' of stars that were already diffuse light in infalling groups and so they become ICL of the host cluster \citep{Rudick2006,Mihos2017}, the disruption of dwarf galaxies \citep{Purcell2007} and `in-situ' stars that formed directly as diffuse light \citep{Puchwein2010}.

Observational studies demonstrate clear gradients in colour and metallicity \citep{DeMaio2015,DeMaio2018,Chen2022,Zhang2024}, which suggest that tidal stripping is the dominant process shaping the ICL at redshift $z=0$. 
Some works also find flat radial profiles, specially in the innermost regions of clusters \citep{Yoo2021,MontesTrujillo2022,Golden-Marx2023}, supporting the idea of a combined formation scenario for the ICL, with the inner regions being built through major mergers and the outer regions growing via tidal stripping of satellites. 
This scenario is generally supported by simulations, which highlight the dominance of stripping \citep[e.g.][]{Contini2019}, but allowing for a contribution from mergers that is more relevant in the inner regions and that can vary widely from one cluster to another depending on properties such as the cluster's mass or dynamical state \citep{Tang2023,Chun2023,Chun2024,Contini2023b}.

In this context, another debated topic is the mass of the galaxies that contribute their stars to the ICL. The general picture from observations is that the stars are mainly stripped from galaxies with stellar masses around $5 \cdot 10^{10}$ M$_\odot$ \citep{MontesTrujillo2014,MontesTrujillo2018,MontesTrujillo2022,Morishita2017,DeMaio2018}. From the side of simulations, many theoretical studies support this picture, predicting galaxies with masses between $10^{10}$ and $10^{11}$ M$_\odot$ to be the main contributors of stars to the ICL \citep{Contini2014,Contini2019,Chun2023,Chun2024,Ahvazi2024progs,Brown2024}. As before, the situation can be very different for individual clusters, with more massive clusters having a more significant contribution from more massive satellite galaxies \citep{Contini2023b,Chun2024}. Nevertheless, the debate remains open, and other simulation studies such as \citet{Tang2023} suggest that the main source of ICL stars are intermediate mass galaxies (with stellar mass from $10^{8}$ to $10^{10}$ M$_\odot$).

Regarding the other previously mentioned formation channels for the ICL, apart from merging and stripping of galaxies, the contribution from totally disrupted dwarf galaxies has been shown to be minor \citep[e.g.][]{Murante2007,DeMaio2018}. As for the accretion of stars that were already diffuse light in another group or halo, the contribution from this channel is unclear, but some works point to a relevant contribution \citep[e.g.][]{Ragusa2023} that becomes more significant for more massive and dynamically unrelaxed clusters \citep{Contini2023b,Chun2024,Brown2024}. Finally, the presence of `in-situ' stars, formed directly as ICL, is a subject of ongoing discussion within the community. While some observational studies find that their contribution is negligible \citep{Melnick2012}, others report a more significant role for this channel, proposing that around 15 to 20 per cent of the ICL is formed in this way \citep{Hlavacek-Larrondo2020,Barfety2022}. Hydrodynamical simulations also show disagreement between them, with some predicting a significant fraction \citep[][for IllustrisTNG]{Montenegro-Taborda2023,Ahvazi2024insitu} and others reporting fractions below 1 per cent \citep[][for \textsc{Horizon-AGN}]{Brown2024}. 

The aim of this paper is to address all these open questions through a set of hydrodynamical simulations of galaxy clusters. In our previous, introductory study \citep{Contreras-Santos2024}, we presented and characterised the ICL in \textsc{The Three Hundred} simulations. This is a set of 324 regions of radius 15 $\hMpc$ centred on the most massive objects from a parent cosmological dark matter only simulation (1 $\hGpc$ side length). These regions were re-simulated with two different hydrodynamical codes: \gx, a smoothed particle hydrodynamics (SPH) solver, and \textsc{Gizmo-Simba}, a meshless hydrodynamic and gravity solver \citep{Hopkins2014,Hopkins2017}. In \citet{Contreras-Santos2024} we presented the properties of the ICL of the 324 clusters at $z=0$, as well as its relation to the dark matter component of the clusters. In the present work, we extend this analysis to higher redshifts by selecting all ICL particles at $z=0$ and tracing them back in time through the different snapshots. This allows us to explore the origin and assembly of the ICL, addressing questions about when, where and how it formed within galaxy clusters.

This paper is organised as follows. In \Sec{sec:data}, we present \textsc{The Three Hundred} simulations data set, the two different hydrodynamical codes, the identification of galaxies and dark matter halos, and our definition of ICL in the simulations. In \Sec{sec:icl-formation}, we present our results regarding the formation of the ICL in the simulated clusters. We first address the timescales of ICL formation and then we discuss the different ways the stars can be contributed to the ICL. In \Sec{sec:strippedICL}, we focus on one of the channels, namely the particles that are stripped from other halos to the ICL, and investigate the properties of these halos. Finally, we summarise and discuss the main results of this work in \Sec{sec:conclusions}.

\section{The Data} \label{sec:data}

\subsection{The Three Hundred sample} \label{sec:data:300}
\textsc{The Three Hundred} project \citep{Cui2018} is an international collaboration aimed at understanding the formation and evolution of massive galaxy clusters. The project data set includes 324 galaxy clusters, each with a mass exceeding $\sim 8 \cdot 10^{14} \hMsun$, drawn from the MultiDark Planck 2 (MDPL2) dark matter-only simulation \citep{Klypin2016}. This simulation traces the hierarchical assembly of $3840^3$ dark matter particles in a (1 $\hGpc$)$^3$ volume, with a particle mass resolution of $1.5\cdot 10^9$ $\hMsun$ and softening length of 6.5 $\hkpc$. It adopts a $\Lambda$CDM cosmology consistent with Planck 2015 \citep{Planck2015}. The 324 most massive halos in MDPL2, identified by the \textsc{rockstar} halo finder \citep{Behroozi2013}, form the core sample of this project.

To run hydrodynamical simulations of these clusters, the regions of interest were refined using the \textsc{Ginnungagap2} code with a zoom-in approach. This method preserves high resolution within a 15 $\hMpc$ sphere around each cluster, while lowering the resolution outside this region. This setup allows computational focus on the cluster environment while retaining the large-scale gravitational field of the simulation box. Dark matter particles in this sphere were split into dark matter and baryonic components based on the $\Omega_\mathrm{m} / \Omega_\mathrm{b}$ ratio, yielding particle masses of $m_{\mathrm{DM}}=1.27 \cdot 10^9$ $\hMsun$ and $m_{\mathrm{gas}}=2.36 \cdot 10^8$ $\hMsun$. Each selected region was then re-simulated using the \textsc{Gadget-X} and \textsc{Gizmo-Simba} codes.

\subsection{Hydrodynamical models} \label{sec:data:models}

In this paper we work with two different implementations of the hydrodynamics adopted in \textsc{The Three Hundred} galaxy clusters, \textsc{Gadget-X} and \textsc{Gizmo-Simba}. The simulated regions share the same initial conditions in both data sets, which allows us to perform code-to-code comparisons on the same clusters. Here we will only briefly mention the main features of each of the two codes. For a more in depth description, including all the necessary details, the reader is referred to the introductory papers by \citet{Cui2018} and \citet{Cui2022}, respectively \citep[see also Section 2.2 in][for a detailed description of both implementations]{Contreras-Santos2024}.

\textsc{Gadget-X} is an improved version of the non-public \textsc{Gadget-3} code \citep{Murante2010,Rasia2015,Biffi2017,Planelles2017}, a tree particle mesh (TreePM) SPH code\footnote{See \citet{Springel2005} for a detailed description of the last public version \textsc{Gadget-2}}. It utilises a density-entropy formulation with a Wendland $C^4$ kernel to improve the accuracy of hydrodynamic calculations \citep{Beck2016,Sembolini2016}. It includes gas cooling from \citet{Wiersma2009} and star formation following \citet{Tornatore2007}, which in turn incorporates the \citet{SpringelHernquist2003} algorithm. Chemical enrichment tracks 11 elements produced by Type II and Type Ia supernovae and AGB stars. Type II supernovae are the only contributor to kinetic stellar feedback, which follows the prescription of \citet{SpringelHernquist2003}. Black hole growth and active galactic nucleus (AGN) feedback are implemented following \citet{Steinborn2015}. Black holes are introduced as collisionless particles within sufficiently massive halos, with growth following the Bondi-Hoyle-Lyttleton \citep{Bondi1952} accretion scheme capped at the Eddington limit. 

The \textsc{Gizmo-Simba} simulations in \textsc{The Three Hundred} project use the \textsc{Gizmo} code \citep{Hopkins2015,Hopkins2017}, which combines SPH with meshless finite mass (MFM) and volume (MFV) methods to solve hydrodynamics, together with the same galaxy formation model as in the \textsc{Simba} simulations \citep{Dave2019}. The \textsc{Simba} model, incorporated into \textsc{Gizmo}, includes advanced galaxy formation subgrid models, building on its predecessor \textsc{Mufasa} \citep{Dave2016}. Gas cooling is implemented using the \textsc{Grackle-3.1} library \citep{Smith2017}, while star formation is modelled using an $\mathrm{H_2}$-based approach following \citet{KrumholzGnedin2011}.
Stellar feedback in \textsc{Gizmo-Simba} includes kinetic outflows driven by Type II and Type Ia supernovae, as well as AGB stars, following the two-phase wind model of \citet{Dave2016}. These winds carry metals that are traced for 11 chemical elements. Black holes are seeded in galaxies with masses above $10^{10.5}$ $\Msun$ and grow via a two-mode accretion model, where hot gas follows Bondi-Hoyle-Lyttleton accretion and cold gas follows a torque-limited model \citep{HopkinsQuataert2011}. The model includes two forms of black hole feedback: kinetic outflows and feedback from X-ray sources, the latter helping to quench the most massive galaxies.

\subsection{Halo catalogues and merger trees} \label{sec:data:ahf}
To identify dark matter halos in the simulations we use Amiga Halo Finder (AHF\footnote{\url{http://popia.ft.uam.es/AHF}}, \citealp{Gill2014,KnollmannKnebe2009}), using the same halo catalogues as previous studies from \textsc{The Three Hundred} collaboration. AHF locates density peaks in the simulation volume, similarly to adaptive mesh refinement (AMR) techniques, by iteratively dividing the space where the total matter density exceeds a threshold. Particles of all species (dark matter, stars, gas, and black holes) are then grouped into gravitationally bound halos. These halos are defined as spherical regions with at least 20 particles and a mass density of $\Delta \rho_\mathrm{crit} (z)$, where $\rho_\mathrm{crit}$ is the critical density of the Universe and $\Delta$ is the corresponding overdensity. In this work, we adopt a fixed overdensity value of 200. This allows the computation of the corresponding masses and radii for the halos, $M_{200}$ and $R_{200}$. 
As for the substructure, subhalos in AHF are identified as smaller halos that lie within the radius of a more massive host halo. Their masses are computed as the total mass gravitationally bound to their centres, rather than using an overdensity threshold, due to their embedding in the host’s density field \citep[see the Appendix of][for more details]{KnollmannKnebe2009}. 

To follow the evolution of the identified halos with redshift, we need to trace them through the different snapshots. For this purpose, we use merger trees created with the \textsc{MergerTree} tool, that comes with the AHF package. \textsc{MergerTree} follows each halo identified at $z=0$ backwards in time, using a merit function to identify its progenitors in the previous snapshot. The main progenitor of halo A is the halo B (at a previous redshift) that maximizes the merit function: $\mathcal{M}=N^2_{AB}/(N_A \cdot N_B)$, where $N_A$ and $N_B$ are the number of particles in halos A and B, respectively, and $N_{AB}$ is the number of particles that are in both halos A and B. In practice, this means that the main progenitor of halo A is not necessarily the most massive one, but rather the one that shares the largest fraction of particles with it. Another essential feature of \textsc{MergerTree}, which corrects for halo finder incompleteness, is that it allows for skipping snapshots if no suitable progenitor is found in the immediately preceding snapshot \citep{Wang16}. In this way, we define the main branch of a halo as the complete list of its main progenitors throughout its cosmic history. For more details on the performance of \textsc{MergerTree} (and also different treebuilders) see \citet{Srisawat13}.

\subsection{ICL definition} \label{sec:data:icl}

In order to study the ICL, the first step is to have a clear operational definition, particularly to separate it from the BCG, where the boundary is physically ambiguous. Given the variety of approaches in the literature, we here outline and motivate the definition adopted in this work.

In this study, we work with the 324 galaxy clusters located at the centres of the resimulated regions. Following \citet[][Section~3]{Contreras-Santos2024}, we define the BCG as all stellar particles within a fixed 50 kpc spherical aperture. We chose this radius for consistency with other works, but we have previously tested that changing to 30 and 70 kpc only shows minor differences. The fixed aperture method has often been adopted in simulations \citep[e.g.][]{McCarthy2010,Ragone-Figueroa2018,Pillepich2018b}, supported by theoretical arguments that advocate arbitrary thresholds due to the absence of a clear physical boundary between the BCG and ICL \citep{Pillepich2018b}. Observations of massive clusters also indicate that this method is adequate for identifying their BCGs \citep{Stott2010,Kravtsov2018}. Although other approaches exist, such as combining BCG and ICL into a single component \citep[e.g.][]{DeMaio2018,Zhang2019} or identifying a transition radius where the diffuse component becomes dominant \citep[e.g.][]{Contini2022,Proctor2023}, we prefer the simplicity and reproducibility of a fixed aperture. 

Once the BCG is defined, we use the AHF halo catalogues to remove stellar particles bound to subhalos. By excluding both BCG and subhalo-associated particles from the total stellar content of each cluster\footnote{The total stellar content of a cluster refers to all bound particles within $R_{200}$.}, the remaining stars define the ICL, a smooth, diffuse component surrounding the more concentrated mass of the BCG.

We emphasise that the 50 kpc aperture definition of the BCG is applied exclusively to the 324 central clusters at $z=0$, where the impact of this choice has been carefully evaluated. Extending this definition to lower-mass systems would require a more detailed analysis and likely further adjustments to account for their different structural properties. Consequently, for subhalos, we treat their stellar content as a whole and exclude it entirely from the main cluster's ICL. This may lead to an underestimation of the ICL, as massive subhalos can host a non-negligible amount of diffuse stars not counted as ICL. Nevertheless, \citet{Contreras-Santos2024} showed that our BCG and ICL mass estimates, although slightly higher at the centre, remain broadly consistent with observations (see their Fig.~3).

Similarly, to ensure consistency and minimise potential biases, we refrain from applying our ICL definition at higher redshifts. Thus, when tracing the history of ICL particles at $z>0$, we consider them to be part of the ICL as long as they are bound to the main cluster halo and not to any other subhalo, even if they are within 50 kpc of the cluster centre. This allows for a consistent and well-defined analysis of the ICL assembly history, which we describe in the next section.

\section{ICL formation} \label{sec:icl-formation}

In this section, we investigate the formation of the ICL by identifying its constituent stellar particles at $z=0$ and tracing their unique IDs backward in time through the simulation snapshots. This allows us to determine when (\Sec{sec:when}), where (\Sec{sec:where}), and how (\Sec{sec:how}) they became part of the ICL, building a comprehensive picture of its origin. 

\subsection{When? Assembly vs formation} \label{sec:when}

Following earlier works such as \citet{DeLucia2007} and \citet{Montenegro-Taborda2023}, we distinguish between the formation time of stars -- the time when they were formed -- and the assembly time -- the time when the star particle first enters the ICL. Specifically, we define a star particle's assembly time as the first time it appears  in one of the main progenitors of the central halo (i.e. in the main branch, see \Sec{sec:data:ahf}) outside of any subhalos, and remains in the ICL for at least two consecutive snapshots. This criterion avoids noise from transient classifications (e.g. particles oscillating between components). However, we tested a stricter definition using based only on the first ICL appearance and found no significant difference.

\begin{figure}
\centering
  \hspace*{-0.3cm}
  \includegraphics[width=9cm]{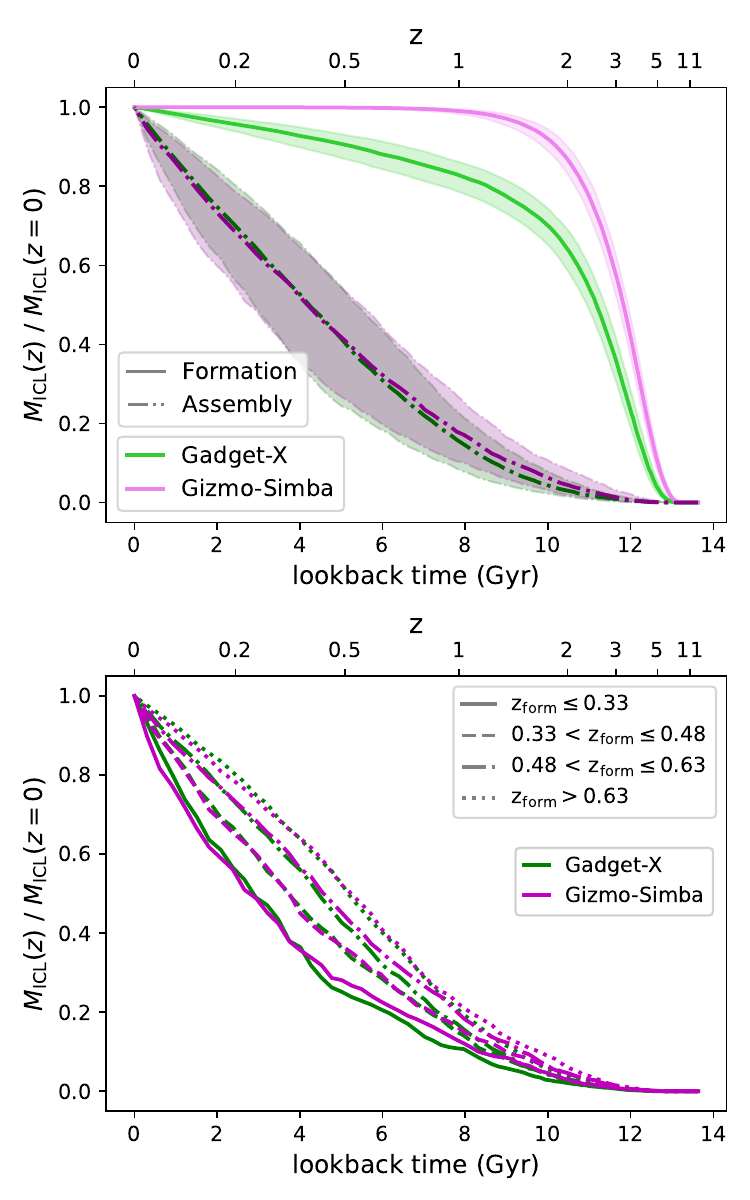}
  \caption{Top, ICL formation (solid lines) and assembly (dash-dotted lines) histories of the 324 clusters for the two different simulation runs, \textsc{Gadget-X} (green) and \textsc{Gizmo-Simba} (magenta). The lines show the median values, with the masses normalized to the total ICL mass at $z=0$, while the shaded regions correspond to the 16th to 84th percentile ranges. Bottom, ICL assembly history separating the clusters into four bins according to their formation time. Bins are selected to have approximately the same number of clusters in each bin.}
  \label{fig:assembly-formation}
\end{figure}

The top panel of \Fig{fig:assembly-formation} shows the formation and assembly times of stars that make up the ICL at $z=0$. We track these times for all ICL particles across the 324 clusters, normalising each history to the cluster's total ICL mass at $z=0$. The figure shows the median formation (solid lines) and assembly (dash-dotted lines) histories, with shaded regions indicating the 16th to 84th percentiles. Green curves correspond to the \textsc{Gadget-X} runs, while magenta curves represent \textsc{Gizmo-Simba}.

In this figure, we can see that the formation histories are significantly different for the two simulations, with \textsc{Gadget-X} showing continuous star formation, even at later times. While for \textsc{Gizmo-Simba} $\sim 99$ per cent of the stars are already formed by $z\sim 1$, in \textsc{Gadget-X} there are only $\sim 80$ per cent of the stars formed by this redshift. This is consistent with Fig. 3 of \citet{Cui2022} that shows a much higher stellar fraction in the \textsc{Gizmo-Simba} clusters at $z>1$, due to the more efficient star formation model followed by a rapid quenching due to intense AGN feedback. For a more detailed discussion, we refer the reader to Section 4 in \citet{Cui2022}.

Regarding the assembly histories, we see that both simulations agree almost perfectly. The assembly of the ICL happens, by definition, much later than the formation of its stars. According to \Fig{fig:assembly-formation}, half of the present day ICL mass was in place between $z\sim 0.2$ and 0.5, while the other half was assembled after that. Going to higher redshifts, for instance $z=1$, we can see that 10 to 30 per cent of the final ICL was already in place by this redshift. We emphasise that these trends represent the assembly history of the $z=0$ population, rather than the actual ICL content at earlier epochs, and are therefore not directly comparable to high redshift observations.
The similarity in the assembly histories is consistent with the fact that both simulations share the same initial conditions and dark matter halo assembly. This highlights that the ICL assembly is largely governed by the hierarchical growth of structure, whereas the differences in formation histories reflect the distinct baryonic physics models used in each simulation.

Given the size of our sample, we can also study how the assembly of the ICL varies from one cluster to another. In the top panel of \Fig{fig:assembly-formation} we can see that the shaded regions for the assembly histories, which correspond to the 16th-84th percentiles, are wider than for the formation histories. This means that, while (future) ICL stars are formed at similar times for all clusters, the time when they are assembled into the ICL changes more significantly from one cluster to another.

The bottom panel of \Fig{fig:assembly-formation} shows the assembly of the ICL, similar to the dash-dotted line in the top panel, but now separating the clusters into four bins according to their cluster formation time, $z_\mathrm{form}$. Following previous works with \textsc{The Three Hundred} simulations \citep[e.g.][]{Mostoghiu2019}, $z_\mathrm{form}$ is defined as the redshift at which $M_{200}$ is equal to half its value at $z=0$. 
The bins in the bottom panel of \Fig{fig:assembly-formation} are selected to ensure approximately the same number of clusters in each bin, for both \textsc{Gadget-X} and \textsc{Gizmo-Simba}. We find a clear trend, such that clusters that formed later (lower values of $\mathrm{z_{form}}$, solid lines), also assemble their ICL later. This points to a tight connection between the formation of the cluster and the build-up of its ICL. Previous works have shown that more dynamically evolved systems -- i.e. those that are earlier assembled, more concentrated, or more relaxed -- tend to host a higher fraction of their stellar mass in the ICL, both in simulations \citep{Rudick2011,Cui2014,Contini2023a,Contreras-Santos2024,Montenegro-Taborda2025} and observations \citep{MontesTrujillo2018,Poliakov2021,Ragusa2023}. Our results reinforce this picture from a theoretical perspective, showing that dynamically older clusters not only have more ICL, but also assemble it earlier.

\subsection{Where? In-situ vs ex-situ} \label{sec:where}

\subsubsection{In-situ fraction}

Now that we know the timescales of ICL formation, the next step is to investigate where these stars formed. We are especially interested in the prevalence of `in-situ' stars, formed directly within the ICL, which means that they already belong to the ICL at the moment of their formation. To identify these, we compare the assembly and formation times of all star particles that end up constituting the ICL at $z=0$. `In-situ' stars are those whose assembly time is equal to the formation time, that is, they were formed directly in the main cluster halo but not in any individual galaxy. We compute the total mass of in-situ stars for each cluster and express it as a fraction of the total ICL mass at $z=0$. 

\begin{figure}
\centering
  \includegraphics[width=8.5cm]{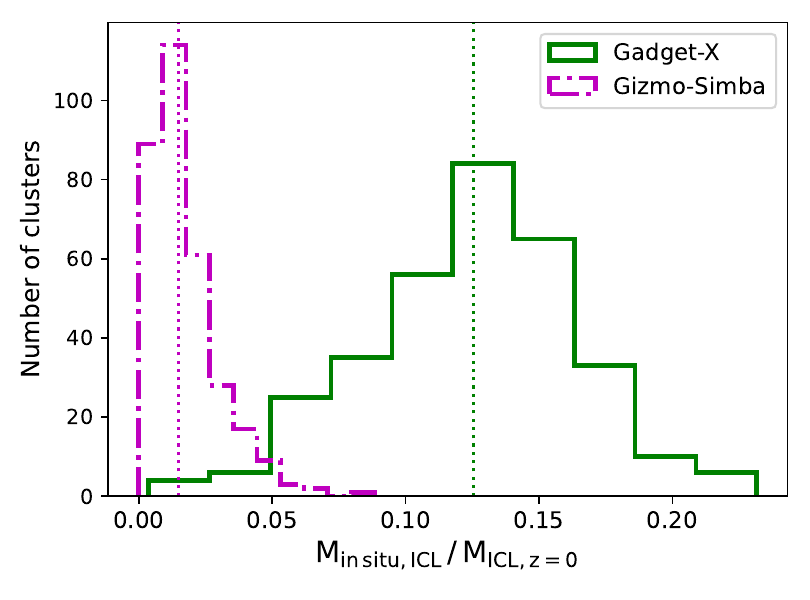}
  \caption{Distribution of the fraction of in-situ ICL mass for the 324 clusters in the \textsc{Gadget-X} (solid green line) and \textsc{Gizmo-Simba} (dash-dotted magenta line) runs.}
  \label{fig:insitu-fraction}
\end{figure}

\Fig{fig:insitu-fraction} shows the distribution of this fraction across the 324 clusters, with solid green lines for \textsc{Gadget-X} and dash-dotted magenta for \textsc{Gizmo-Simba}. Vertical dotted lines indicate the median of each distribution. The figure reveals clear differences between the two simulations sets: \textsc{Gadget-X} produces a broad distribution, with some clusters reaching 20 per cent of in-situ ICL mass and a median value of 12.6 per cent, while \textsc{Gizmo-Simba} yields a narrower distribution with a median value of 1.5 per cent. These contrasting results suggest that there are different mechanisms at work in the simulations.

\subsubsection{Distance to host centre}

\begin{figure*}
\centering
  \includegraphics[width=15.5cm]{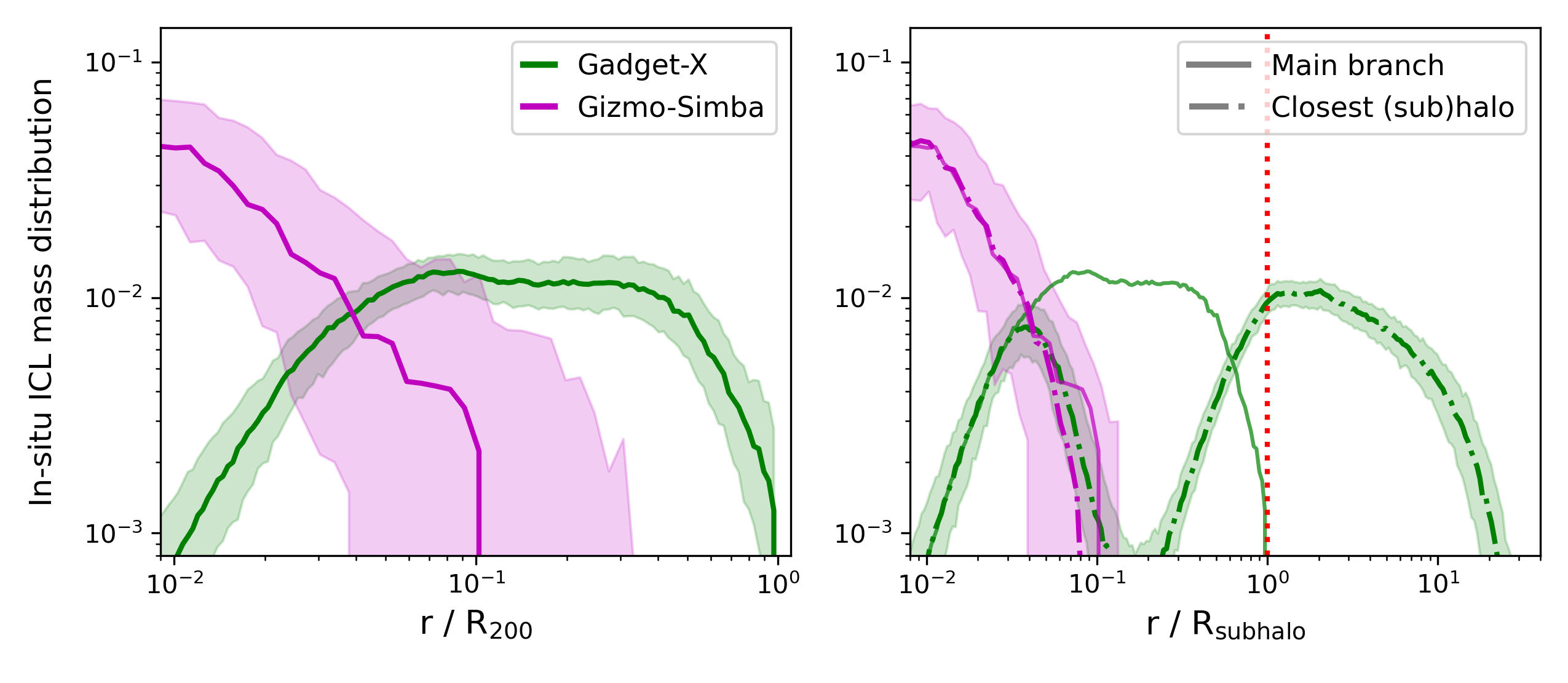}
  \vspace*{-0.4cm}
  \caption{Left, distribution of distance from the in-situ ICL particles to the main branch progenitor centre at the formation time of the particle, in units of the main branch halo radius $R_{200}$. Green is for \textsc{Gadget-X}, magenta for \textsc{Gizmo-Simba}. Right, distance from the in-situ particles to the centre of the closest halo (or subhalo), in units of the radius of that halo, $R_\mathrm{subhalo}$. To ease comparison, we still show the distance to the main branch with the corresponding solid lines. Lines show the median computed stacking the 324 clusters, shaded regions are the 16th-84th percentiles.}
  \label{fig:insitu-distance}
\end{figure*}

Given the significant in-situ ICL fractions for \textsc{Gadget-X} clusters, we now explore more closely the formation of these stars. 
First, our definition for in-situ requires that, when the star is formed, it belongs only to the main cluster halo, and not to any subhalo. We are not imposing any criteria in the distance from the centre at which the star is formed, so it could be the case that the star is formed very close to the centre, i.e. in the BCG, and then moved outwards so that at $z=0$ it belongs to the ICL \citep[see, e.g., Section 3.3 in][]{Ragone-Figueroa2018}. To check for this situation, for the stars classified as in-situ, we compute their distance to the main cluster centre at the time they were formed. In the left panel of \Fig{fig:insitu-distance} we show this distance, normalised by the radius $R_{200}$ of the main cluster at the formation time of the star. We compute the distribution of the distances, weighted by particle mass, for each of the clusters, and then in \Fig{fig:insitu-distance} we show the median values for the set of the 324 clusters (solid lines), as well as the 16th to 84th percentiles as shaded regions. For \textsc{Gizmo-Simba} (depicted in magenta), we see that the distribution shows a prominent peak at $r \sim 0$, indicating that the in-situ star formation can be in fact associated with the central galaxy in the cluster. In contrast, \textsc{Gadget-X} (depicted in green) also shows a substantial population of stars at $r \lesssim 0.1 \cdot R_{200}$, but with an extended distribution reaching up to $\sim 0.5 \cdot R_{200}$. This suggests that a notable fraction of stars formed far from the cluster centre but not bound to any substructure at that time or later.

For further investigation into this scenario, we now consider the distance to the closest subhalo at the formation time of the in-situ star particle. In this way, we want to explore the possibility that these in-situ particles not associated with the central galaxy are still associated to another galaxy, even if they are not identified as bound to it by the halo finder.
To obtain this quantity, we compute the distance (in kpc) from each particle to all subhalos inside the cluster in the snapshot it was formed. We select the minimum distance in kpc and the subhalo associated to it and then we normalise by the radius of this subhalo. 
We note that the main cluster halo is included in this calculation, so it can also be selected as the closest substructure, although we do not specify this in the figure notation for simplicity.
In the right panel of \Fig{fig:insitu-distance}, similarly to the left one, we show the distribution of this distance, in units of $R_\mathrm{subhalo}$. As before, the situation for \textsc{Gizmo-Simba} is what could be expected, the particles are always formed in the centre of the halo they are associated to. However, the scenario is different for \textsc{Gadget-X}, for which, while there is also a peak at $r \lesssim 0.1 \cdot R_{200}$, there is another peak at $r \sim R_\mathrm{subhalo}$. This peak shows tails in both directions of the $x$-axis. To the right, this means that particles are being formed in the outer edges of subhalos and not bound to them. The tail to the left is more surprising, since it represents star particles formed inside subhalos without being bound to them. 

\subsubsection{Interpretation and discussion} \label{sec:where:discussion}
Summarising the previous results, we find two different situations for our two simulations. For \textsc{Gadget-X} we have seen that the fraction of in-situ ICL mass ranges from $\sim 5$ to almost 20 per cent. While most of the in-situ stars are associated with the central galaxy, that is, formed in the centre of the cluster and then moved outwards, there is a significant fraction ($\sim$ 35$\%$) of in-situ stars that are formed at a distance of more than $\sim 0.2 \cdot R_{200}$ from the cluster centre. For \textsc{Gizmo-Simba}, the fraction of in-situ ICL mass peaks at $\sim 1.5$ per cent, and for some clusters it reaches values of 5 per cent. However, all the in-situ star formation is associated with the central galaxy in this case.

To better understand the origin of non-central in-situ stars in \textsc{Gadget-X}, we further examined the subhalo closest to each star at its formation time. While our approach is deliberately approximate and we do not include additional figures, it reveals clear and consistent trends. The reported numbers should be seen as lower limits, since associating objects based on physical distance may occasionally misidentify unrelated structures. 

Analysing the gas content of the subhalos associated with in-situ stars formed at $r > 0.2R_{200}$, we find that the great majority ($\sim 80\%$) are strongly gas-dominated, with more than 90 per cent of their total mass in gas. When identifying the closest companion to these gas-dominated subhalos, we find that in 70$\%$ of the cases it is a gas-free object, and in the remaining 30$\%$ it is another nearly pure gas blob. Given the nature of the merger trees, we cannot trace back in time such gas-dominated objects, but only those containing dark matter. Therefore, for the 70$\%$ of companions that do contain dark matter (that is, the gas-free ones), we compute their infall time and compare it with the formation time of the associated in-situ ICL particles. We find that, at the time of ICL star formation, 60$\%$ of these companions have been inside the cluster for less than 2 Gyr, while also 60$\%$ of them have a velocity relative to the cluster that exceeds the cluster’s velocity dispersion, thus linking non-central in-situ ICL to fast-moving, recently infalling subhalos.

These findings establish a clear scenario, in which a galaxy –initially composed of dark matter, stars and gas– falls into the cluster and experiences gas stripping. This process results in two distinct objects identified by the halo finder: a gas-dominated subhalo and a companion gas-free subhalo consisting of the leftover stars and dark matter. As the gas blob is decelerated at the cluster entrance, new stars can still form within it. However, since stars are collisionless, they quickly become unbound from the gas blob. In the first snapshot they appear, some of these stars can be already outside the gas blob, while others can remain inside but unbound, explaining the tails in both directions of $r=R_\mathrm{subhalo}$ in the right panel of \Fig{fig:insitu-distance}.

While this mechanism accounts for the in-situ star formation outside the centre of the cluster, the question remains as to why these gas blobs form and whether they should survive or be shredded. 
A deep exploration of this topic would require a much more detailed study, which is out of the scope of this paper. However, we suggest that the gas may be stripped via ram-pressure stripping or artificially disrupted due to the different density regions the objects are going through \citep[see, e.g.,][]{Mostoghiu2021}. The survival of the stripped blobs depends on how well the code deals with Kelvin-Helmholtz (KH) instabilities. SPH codes such as \textsc{Gadget-X} have been shown to struggle in dealing with these instabilities \citep{Agertz2007}, and so this scenario is favoured in \textsc{Gadget-X}.
Additionally, the formation of the gas blobs also depends on the gas reservoir the galaxies retain, which is tied to the strength of galaxy feedback. In \textsc{Gizmo-Simba}, the stronger feedback implemented prevents the infalling galaxies from retaining enough gas for this scenario to occur. 
Together, these two aspects of the simulations --dealing with KH instabilities and galaxy feedback--, indicate that the fraction of non-central in-situ star formation is highly simulation and code dependent and so very uncertain.

Comparing to similar works tracing the ICL particles in hydrodynamical simulations, some differences and similarities can be found. Using the TNG300 cosmological simulation of the IllustrisTNG project,  \citet{Montenegro-Taborda2023} find a mean in-situ ICL fraction of 11.4 per cent for galaxy clusters, close to our \textsc{Gadget-X} fraction. In the TNG50 simulation within the same project, \citet{Ahvazi2024insitu} focus on three objects with masses $\sim 10^{14}$ M$_\odot$ and find an in-situ fraction that goes from 8 to 28 per cent. Computing the distance to the cluster centre, they also find star formation out of the central region (out to 0.8$R_{200}$) that is not associated to any substructure, similarly to our \textsc{Gadget-X} results. However, \citet{Ahvazi2024insitu} further state that this star formation is not associated with ram pressure stripping of gas in infalling galaxies, but rather to small cloud-like objects condensing along filamentary structures that follow the neutral gas and dark matter distribution. Working with the \textsc{eagle} simulations, \citet{Proctor2023} find that, at the cluster scale, around 10 per cent of the ICL mass is created in-situ. In contrast, in a sample of 11 clusters from the \textsc{Horizon-AGN} simulations, \citet{Brown2024} find that the contribution from in-situ star formation to the ICL is negligible, more aligned with our results for \textsc{Gizmo-Simba}. The same is concluded by \citet{Joo2024} using a sample of more than 1200 groups and clusters from the Horizon Run 5 simulation. 

These trends appear broadly consistent with the implications of our proposed mechanism. Simulations based on adaptive mesh refinement (AMR, e.g. \textsc{Ramses}), such as \textsc{Horizon-AGN} and Horizon Run 5, are better at capturing fluid instabilities and tend to suppress this kind of star formation, while SPH-based simulations such as \textsc{eagle} present substantial in-situ ICL formation. The case of IllustrisTNG is more ambiguous: despite employing the moving mesh code \textsc{Arepo}, which is expected to better resolve instabilities, it still shows significant in-situ ICL. This may be related to specific aspects of its feedback and star formation models, that allow stripped gas to cool and form stars under certain conditions. Overall, while our proposed scenario offers a possible explanation for these differences, a more definitive assessment would require targeted comparisons and deeper analysis.

\subsection{How? Formation channels} \label{sec:how}

\begin{figure*}[ht]
    \centering
    \hspace*{0.3cm}
    \begin{subfigure}{0.52\textwidth}
        \includegraphics[width=11cm]{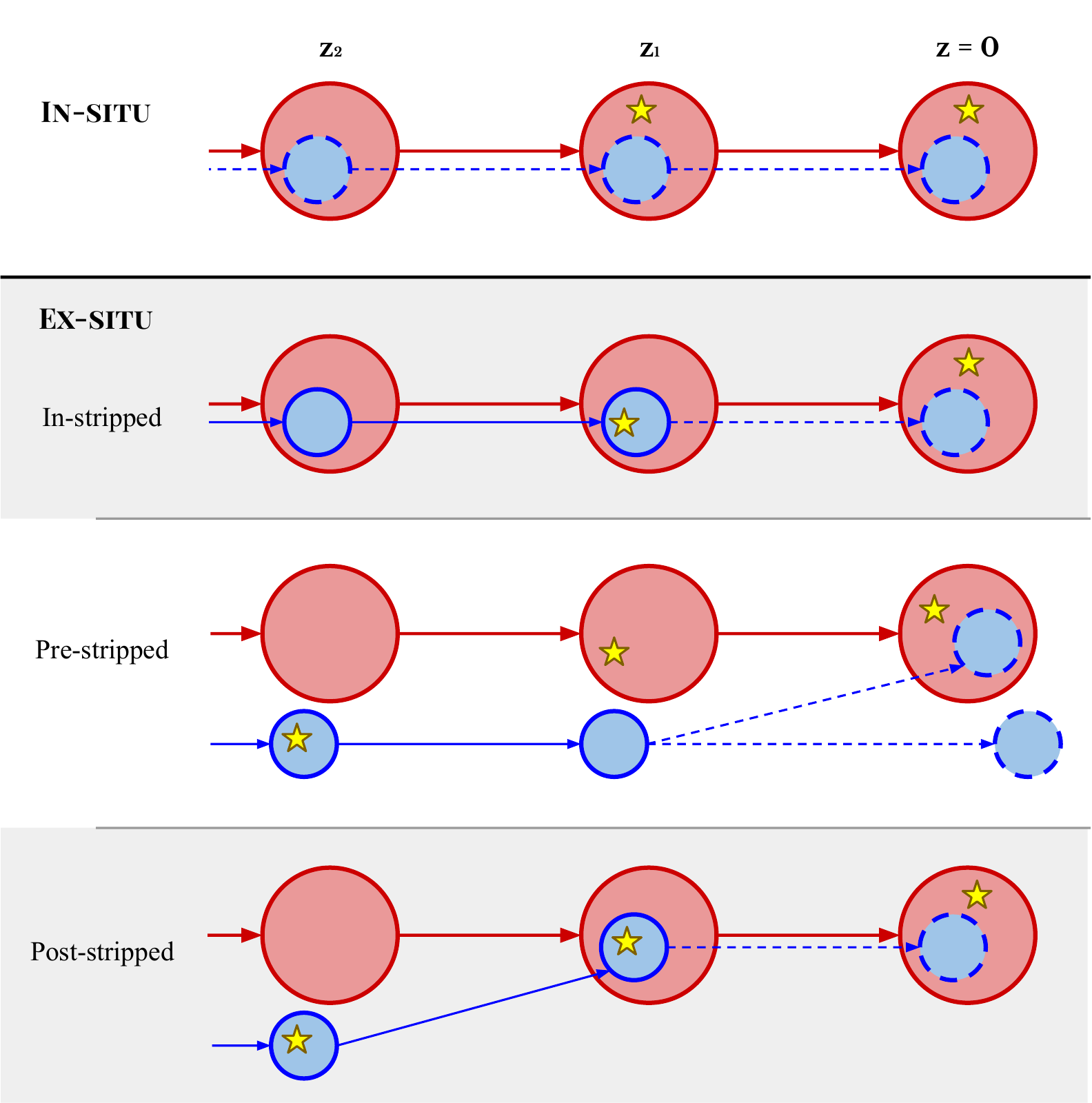}
        \label{fig:plot1}
    \end{subfigure}
    \hfill
    \begin{subfigure}{0.32\textwidth}
        \includegraphics[width=6cm]{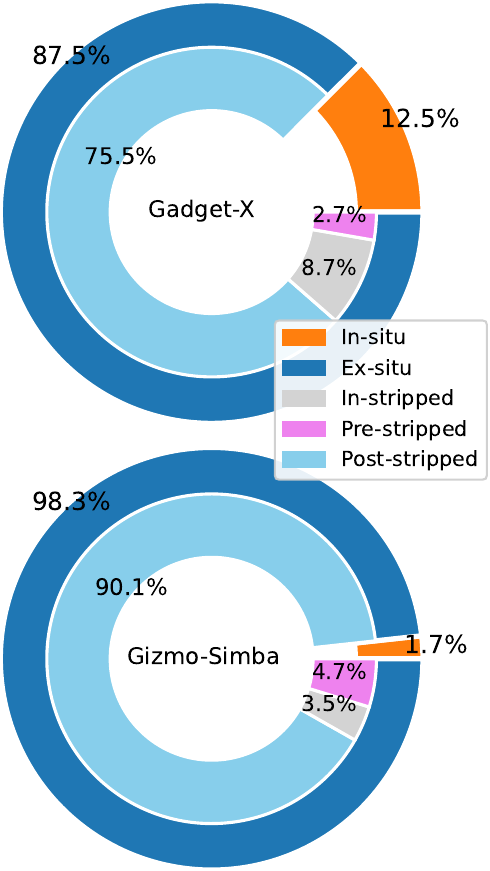}
        \label{fig:plot2}
    \end{subfigure}
    \caption{Classification of the ICL star particles according to the way they were assembled into the ICL. Stellar particles are first classified into in-situ and ex-situ following \Fig{fig:insitu-fraction}, and then ex-situ particles are in turn classified into three different categories. On the left, schematic representation of the different channels, with the main branch depicted in red and a secondary branch in blue. On the right, mean value of the contribution from each channel for the 324 clusters in \textsc{The Three Hundred} sample. Top panel is for \textsc{Gadget-X}, bottom panel for \textsc{Gizmo-Simba}.}    
    \label{fig:stars-classification}
\end{figure*}

In the previous subsection we have classified the ICL stellar particles into ex-situ and in-situ, depending on whether they were formed in a subhalo or directly in the host halo. Following these results (see \Fig{fig:insitu-fraction}), we now summarise this classification in \Fig{fig:stars-classification}. In the right panel of this figure we show, for \textsc{Gadget-X} on top and \textsc{Gizmo-Simba} on the bottom, the ICL mean mass composition of \textsc{The Three Hundred} clusters. The mean in-situ mass fraction is, respectively, 12.5 and 1.7 per cent. For the remaining star particles, the ex-situ ones, we only know that they were formed in a different halo to the main one and at some point they started to be ICL particles. 
To further classify these particles, we focus on the snapshot prior to their assembly into the ICL and identify the halo they belonged to, that is, the halo that contributes them to the ICL. Taking into account the formation and assembly times of each ex-situ star, and comparing with the position of its contributing halo relative to the main cluster halo at these times, we distinguish three different categories. In the three cases, the star has to be stripped from the contributing halo and moved into the main cluster halo, and so we call these three categories `stripped', but including a prefix that denotes the different scenarios. These three categories are the following (see \Fig{fig:stars-classification}, left panel, for a schematic representation of the described processes):
\begin{itemize}
    \item `In-stripped' stars: the particle is formed in a subhalo that is already inside the cluster and is later stripped into the ICL.
    \item `Pre-stripped' stars: the particle is formed in a halo outside the cluster. It is stripped and enters the ICL before its own contributing halo falls into the cluster. We do not distinguish whether these halo afterwards falls inside the cluster or remains outside. The former can happen if the infalling halo `graces' the cluster's outskirts before eventually falling in. The latter can be either because the halo did not have enough time (i.e. the infall will be later than $z=0$) or due to a fly-by, in which the star is torn by the cluster from the outer part of the halo. Since these stars represent a very low percentage we do not believe it necessary to check this scenario in more detail.
    \item `Post-stripped' stars: the particle is formed in a halo outside the cluster, but it is now accreted into the ICL after its contributing halo falls into the cluster. The great majority of ex-situ stars fall within this category.
\end{itemize}

In the right panel of \Fig{fig:stars-classification} we can see that the ICL is widely dominated by the post-stripped stars, stripped from a halo after its infall into the host cluster. If we consider only the ex-situ mass, post-stripped stars represent around 90 per cent for both simulations. For this reason, we will devote the next section to a deeper study of this component of the ICL. 

The second most important contribution comes from stars that are also stripped but that were formed in a halo after its infall into the host cluster. They are more frequent in \textsc{Gadget-X}, which is reasonable given that this simulation has a more significant late star formation (see \Fig{fig:assembly-formation}). We also have to note that, in the \textsc{Gadget-X} run, around $30\%$ of these `in-stripped' stars are formed within a subhalo that is gas-dominated, similarly to the non-central in-situ star formation discussed in \Sec{sec:where:discussion}. In this case, the star is still bound to the gas blob when formed, and so it is not considered in-situ. As before, this scenario does not occur in \textsc{Gizmo-Simba}, where `in-stripped' stars represent a lower fraction of the total ICL, and are all associated to objects that are not gas-dominated. 

\begin{figure*}
\centering
  \includegraphics[width=15cm]{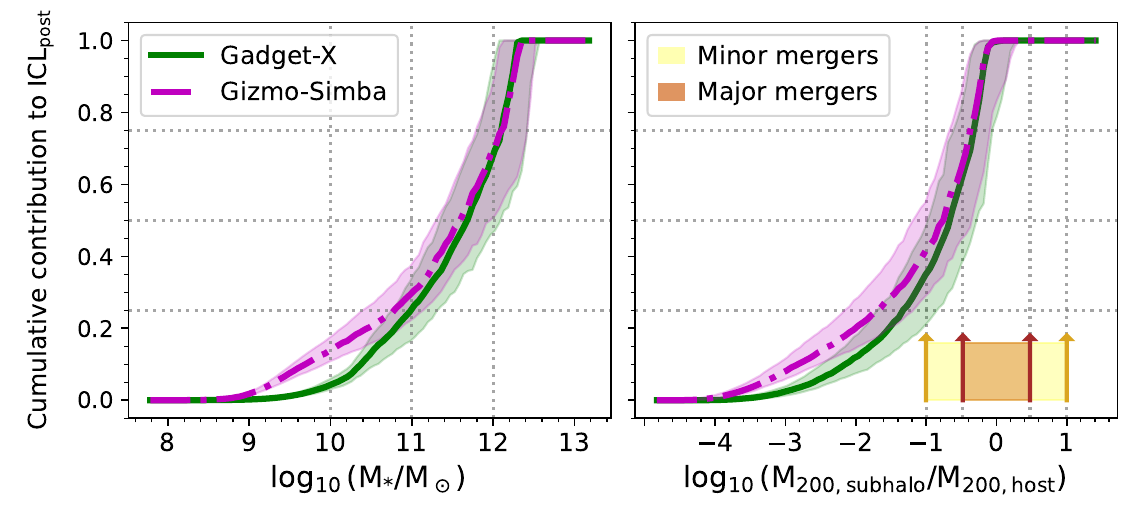}
  \caption{Cumulative distribution of the mass contribution to the post-stripped ICL from halos of different infall masses. Left column is for stellar mass at infall, right column for the total mass ratio to the host at infall time. Lines depict the distributions obtained as the median of the individual distributions for each cluster, while the shaded regions indicate the 16th-84th percentiles. Green solid lines are for \textsc{Gadget-X} and magenta dash-dotted lines for \textsc{Gizmo-Simba}. In the right column, yellow and brown shaded regions indicate the minor (1:10 to 1:3) and major merger regimes (1:3 to 1:1). Horizontal dotted lines indicate fractions of 0.25, 0.5 and 0.75 of the post-stripped ICL.}
  \label{fig:relcontrib-cum}
\end{figure*}

We want to highlight that our distinction between `pre-' and `post-stripped' stars only depends on whether the star is contributed to the ICL by a halo before or after its infall into the host halo. The contributing halo can be a small subhalo or a large halo merging with the central one. We thus do not check if the particle was previously in a galaxy or already ICL of a merging cluster. For instance, if a particle belongs to the diffuse component (intragroup light, IGL) of an infalling group, and the group crosses $R_{200}$ of the host halo before the particle is accreted into the host cluster's ICL, this star would be classified as `post-stripped'. Determining whether stars were already ICL in their original halos requires further study, which we leave for future work. Similarly, we do not consider here the fate of the contributing halos after their particles are stripped. A more detailed analysis is required to determine whether they merge with the BCG, survive as subhalos in the cluster or are totally disrupted, that we also leave for a follow-up work.

\section{Post-stripped ICL} \label{sec:strippedICL}

As discussed in the previous section, the majority of the ex-situ ICL stars are classified as `post-stripped', that is, stars that were stripped from infalling halos after they crossed into the host cluster's $R_{200}$. In this section, we focus on this dominant component by analysing the halos that contribute these stars and their role in building up the ICL.

\subsection{Properties of contributing halos}

We have previously obtained the information about which halos bring the star particles into the ICL. We now examine the properties of these halos. To enable consistent comparisons between different halos, we consider their properties at their infall time, defined as the last snapshot in which the halo is outside $R_{200}$ of the host cluster. For halos that have more than one infall, we only consider the first one. This way, we ensure that the halos are mostly unaltered by their infall into the cluster, and so their properties remain unbiased. 

Throughout this section, we will only work with the `post-stripped' stars (see \Sec{sec:how}) and use the term ICL$_\mathrm{post}$ to refer to this component. We exclude the `in-stripped' and `pre-stripped' categories to ensure consistency and to highlight distinct pathways of ICL formation. In-stripped stars form from the remaining gas of a subhalo already inside the cluster, while pre-stripped stars originate from a halo that may never enter the cluster, making their infall time ill-defined. By concentrating on the dominant post-stripped population, we ensure a clear and consistent analysis of this accretion process.

In \Fig{fig:relcontrib-cum} we show the distribution of the relative contribution to the post-stripped ICL from the different halos. We consider the ICL$_\mathrm{post}$ at $z=0$ of each of the 324 clusters and, for each star particle, we get the mass at (first) infall of its contributing halo, both stellar and total (including also dark matter and gas). The left column of \Fig{fig:relcontrib-cum} focuses on the stellar mass distribution. For each cluster, we compute the cumulative contribution of infalling stellar masses, and then derive the median across all clusters (solid green lines for \textsc{Gadget-X}, dash-dotted magenta for \textsc{Gizmo-Simba}), with shaded regions indicating the 16th-84th percentiles. Horizontal dotted lines mark the 25, 50 and 75 per cent cumulative contribution levels.

At the low-mass end, both simulations show a tail of contributing halos that extends down to the resolution limit --mean stellar particle mass is $\sim 0.6$ ($\sim 2) \cdot 10^8$ M$_\odot$ for \textsc{Gadget-X} (\textsc{Gizmo-Simba}). While the contribution from these low-mass halos is small, \textsc{Gizmo-Simba} predicts a slightly higher relative importance compared to \textsc{Gadget-X}. Despite some incompleteness due to resolution, this is unlikely to strongly affect our results. Higher resolution hydrodynamical simulations (e.g. \citealp{Brown2024}, with stellar mass resolution of $\sim 2 \cdot 10^6$ M$_\odot$; or \citealp{Ahvazi2024progs}) have shown that halos with stellar mass $\lesssim 10^9$ M$_\odot$ contribute less than 10 per cent of the ICL mass. 
Moving to higher masses, the left panel of \Fig{fig:relcontrib-cum} shows that half of the post-stripped ICL is contributed by halos with a stellar mass above $10^{11.4-12.1}$ M$_\odot$, with the distribution extending up to $\sim 10^{12.5}$ M$\odot$. Both simulations show excellent agreement in this regime. These high stellar masses correspond to the stellar component of massive groups and clusters, indicating that there are major mergers involved in the formation of the ICL.

We study this in the right panel of \Fig{fig:relcontrib-cum}, in which we follow the same procedure to plot the cumulative distribution of the total mass ratio between the host and the contributing halo at its infall time. As before, the distribution is skewed left, with a tail that reaches values of the mass ratio smaller than $10^{-4}$. The shaded yellow and brown regions indicate the typical mass ratio ranges associated with minor (1:10 to 1:3) and major (1:3 to 1:1) mergers \citep[e.g.][]{Planelles2009,Contreras-Santos2022a}. Infalling objects with a mass ratio below 1:10 are generally considered as very minor mergers or smooth accretion. These low-mass contributors account for roughly 20–50 per cent of the ICL$_\mathrm{post}$ mass in \textsc{Gadget-X}, and 30–55 per cent in \textsc{Gizmo-Simba}, which again shows a slightly larger contribution from lower-mass halos. The remaining mass is brought in through minor and major mergers, which contribute medians of around 25 and 35 per cent, respectively, though with considerable scatter among clusters, as shown by the shaded regions. The plot also shows a small contribution from mergers with a ratio $>1$, which indicates that the merging object is larger than the one considered the host. This is due to the way the merger trees are built, which is such that the main progenitor does not always have to be the most massive progenitor (see \Sec{sec:data:ahf}). In this way, it can be the case that a bigger halo merges with the main branch halo, bringing some of its mass into the host's ICL. Although not very often, this situation happens sometimes, and contributes some mass to the formation of the ICL of massive clusters.

In general, \Fig{fig:relcontrib-cum} shows how massive galaxies, groups and even other clusters are responsible for building up the ICL of galaxy clusters as massive as those in \textsc{The Three Hundred} sample. It is important to note that, according to \citet{Martin2024}, the dark matter resolution plays a critical role in accurately capturing the stripping process of galaxies, more so than the stellar resolution. Poorly resolved halos can have their stellar material over-stripped, potentially affecting even relatively massive halos. While this issue is particularly relevant for low-mass objects, which may contribute slightly more than our results suggest, it has less impact on very massive objects, which are well resolved in our simulations. Thus, we emphasize the significant contribution from these massive systems in building the ICL.

\begin{figure}
\centering
  \hspace*{-0.2cm}
  \vspace*{-0.3cm}
  \includegraphics[width=8.3cm]{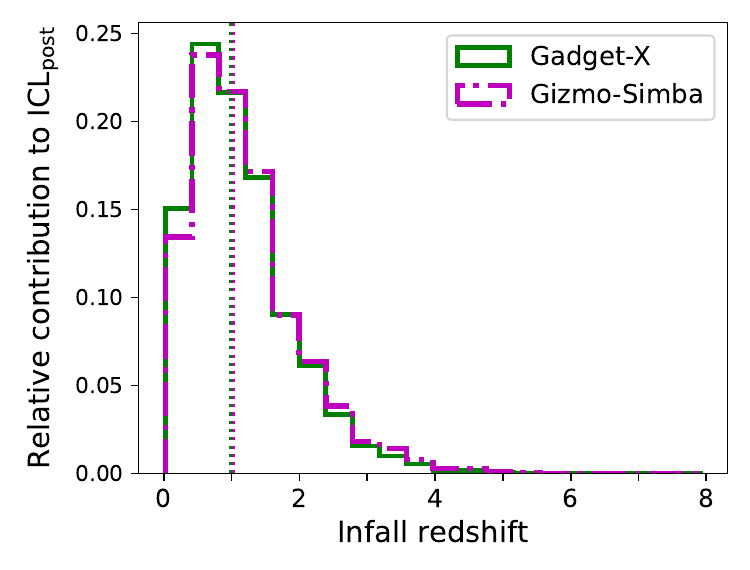}
  \caption{Distribution of the relative mass contribution to the post-stripped ICL from halos with different infall time. The values for the 324 clusters have been stacked together to produce the distributions for \textsc{Gadget-X} (green) and \textsc{Gizmo-Simba} (magenta). The vertical dotted lines indicate the median values of each distribution.}
  \label{fig:relcontrib-zinf}
\end{figure}

To complete this picture of infalling halos that bring the ICL particles into the cluster, \Fig{fig:relcontrib-zinf} shows the distribution of the infall redshift of the contributing halos, computed as the redshift of the last snapshot in which the halo is outside $R_{200}$ of the host cluster. We can see that the maximum values are around $z \sim 4$, while the contributions start to become more significant at redshift around 2. 
This is consistent with the assembly history in \Fig{fig:assembly-formation}, which indicates the time when the stars were accreted into the ICL and shows no ICL growth at $z \sim 4$. By definition, `post-stripped' stars have to enter the cluster before becoming ICL, and so the infall redshifts in \Fig{fig:relcontrib-zinf} have to be higher than those at which the ICL starts growing in \Fig{fig:assembly-formation}. Besides, \Fig{fig:relcontrib-zinf} shows that most halos infall at $z \sim 1$. We also remind here that, for halos that have more than one infall, we only consider the first one. 

\subsection{Number of accreted halos}

\begin{figure}
\centering
  \hspace*{-0.4cm}
  \includegraphics[width=8.9cm]{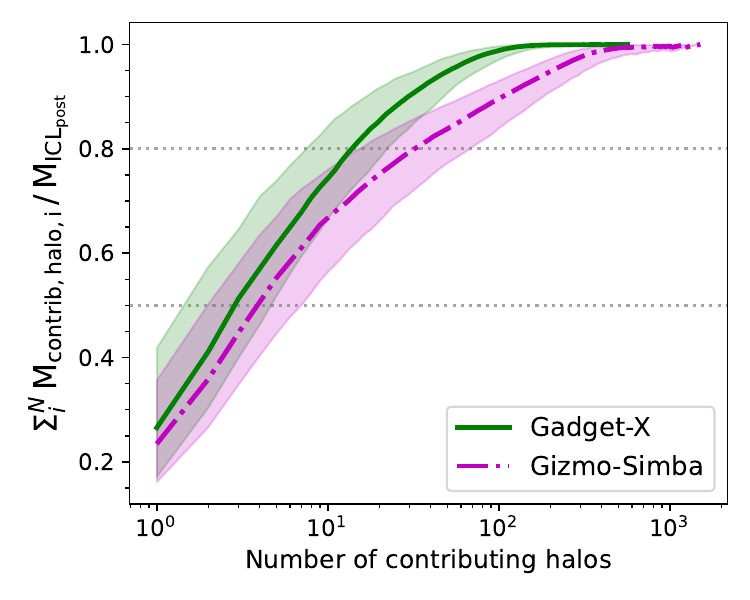}
  \caption{Median number of accreted halos required to assemble the ICL$_\mathrm{post}$. Lines (solid green for \textsc{Gadget-X}, dash-dotted magenta for \textsc{Gizmo-Simba}) indicate the median distribution, while shaded regions denote the 16th-84th percentiles. The halos are ordered starting by the most contributing one, and the values in the $y$-axis are the cumulative contribution of all previous halos, normalised by the ICL$_\mathrm{post}$ mass at $z=0$. Horizontal lines indicate fractions of 0.5 and 0.8 in the $y$-axis, for reference.}
  \label{fig:n-events}
\end{figure}

\begin{figure*}[h]
\centering
  \includegraphics[width=15.7cm]{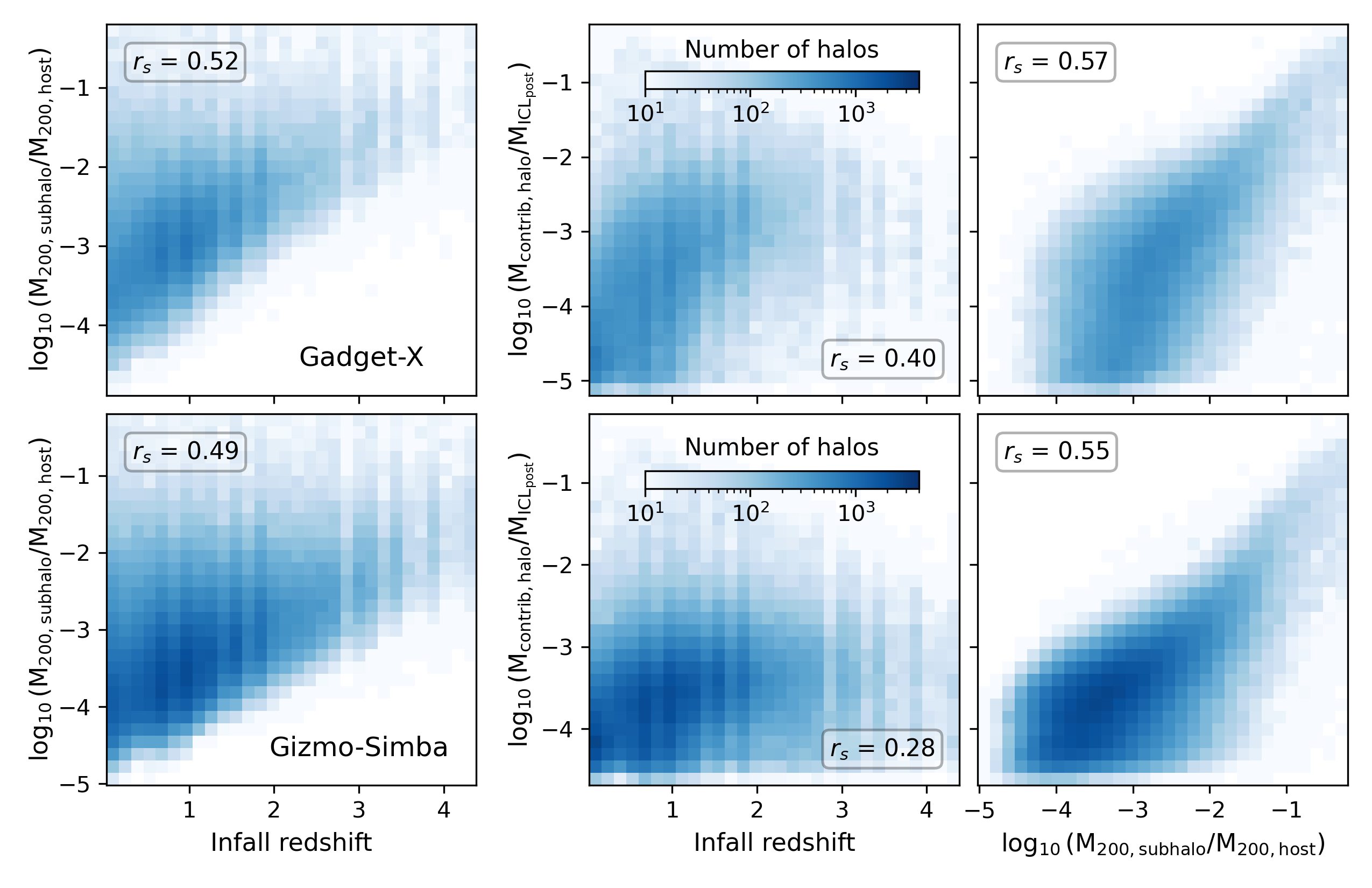}
  \caption{2D histograms showing the distribution of the events that build the ICL$_\mathrm{post}$, top row is for \textsc{Gadget-X}, bottom one for \textsc{Gizmo-Simba}. First column shows the scatter plot between infall redshift and mass ratio to the host cluster, coloured by the number of accreted halos. The second and third columns show the total mass that each halo contributed to the ICL$_\mathrm{post}$ of the involved cluster, as a function of infall redshift and mass ratio of the event. Spearman correlation coefficients, $r_s$, are shown in each panel.}
  \label{fig:contrib-2dhist}
\end{figure*}

We have already selected all the halos that contribute to construct the post-stripped ICL of the simulated clusters, and we have studied the mass, mass ratio and infall time of these objects. Now, we can focus on the contribution that each halo makes to the ICL$_\mathrm{post}$ of the cluster. For each of the contributing halos, we sum up the mass of all the particles that it contributes at any time, and we normalise this mass by the ICL$_\mathrm{post}$ mass at $z=0$ of the cluster. Then, we order the contributing halos by this value (halos that contribute the most go first) and create a cumulative distribution, showing the summed ICL$_\mathrm{post}$ fraction as a number of the involved halos. We repeat this procedure for each of the 324 clusters, and in \Fig{fig:n-events} we show the median distribution together with 16th to 84th percentiles (shaded regions). Essentially, this figure is showing the minimum number of accreted halos needed for a median cluster to assemble a given fraction of its ICL$_\mathrm{post}$. We can see in \Fig{fig:n-events} that already the most contributing halo can reach up to 40 per cent of the final ICL$_\mathrm{post}$ mass. To reach 50 per cent of their ICL$_\mathrm{post}$ mass (indicated with a horizontal line), clusters need a minimum of 2-8 accreted objects for both simulations. The situation starts to differ for them at higher values. For instance, 80 per cent of the ICL$_\mathrm{post}$ is built with 8 to 22 halos for \textsc{Gadget-X}, while \textsc{Gizmo-Simba} requires a minimum number between 15 and 70. For the whole post-stripped ICL at $z=0$\footnote{We compute this value using 99 per cent of the ICL$_\mathrm{post}$ instead of 100, in order to avoid counting minimal contributions.}, the differences become even larger, with 75 to 150 halos needed for \textsc{Gadget-X} and 300 to 600 for \textsc{Gizmo-Simba}. This is in agreement with the results from \Fig{fig:relcontrib-cum}, which showed that the contributing halos are less massive in \textsc{Gizmo-Simba}.
A possible explanation is that galaxies in \textsc{Gizmo-Simba} develop denser stellar cores than those in \textsc{Gadget-X} \citep{Meneghetti2023}, making them less susceptible to tidal stripping. As a result, larger galaxies in \textsc{Gizmo-Simba} are more likely to retain their identity rather than contributing to the ICL, while smaller satellites, which are more easily disrupted, dominate the ICL build-up. On the other hand, in \textsc{Gadget-X}, lower-mass galaxies are more susceptible to disruption, sometimes even before reaching $R_{200}$, which prevents them from contributing to the ICL. This could explain why \textsc{Gizmo-Simba} has a higher contribution from low-mass halos. See, e.g., \citet{Dolag2009} for a detailed study on the survival rates of infalling substructures in hydrodynamical simulations.

\begin{figure}[h]
\centering
  \includegraphics[width=8cm]{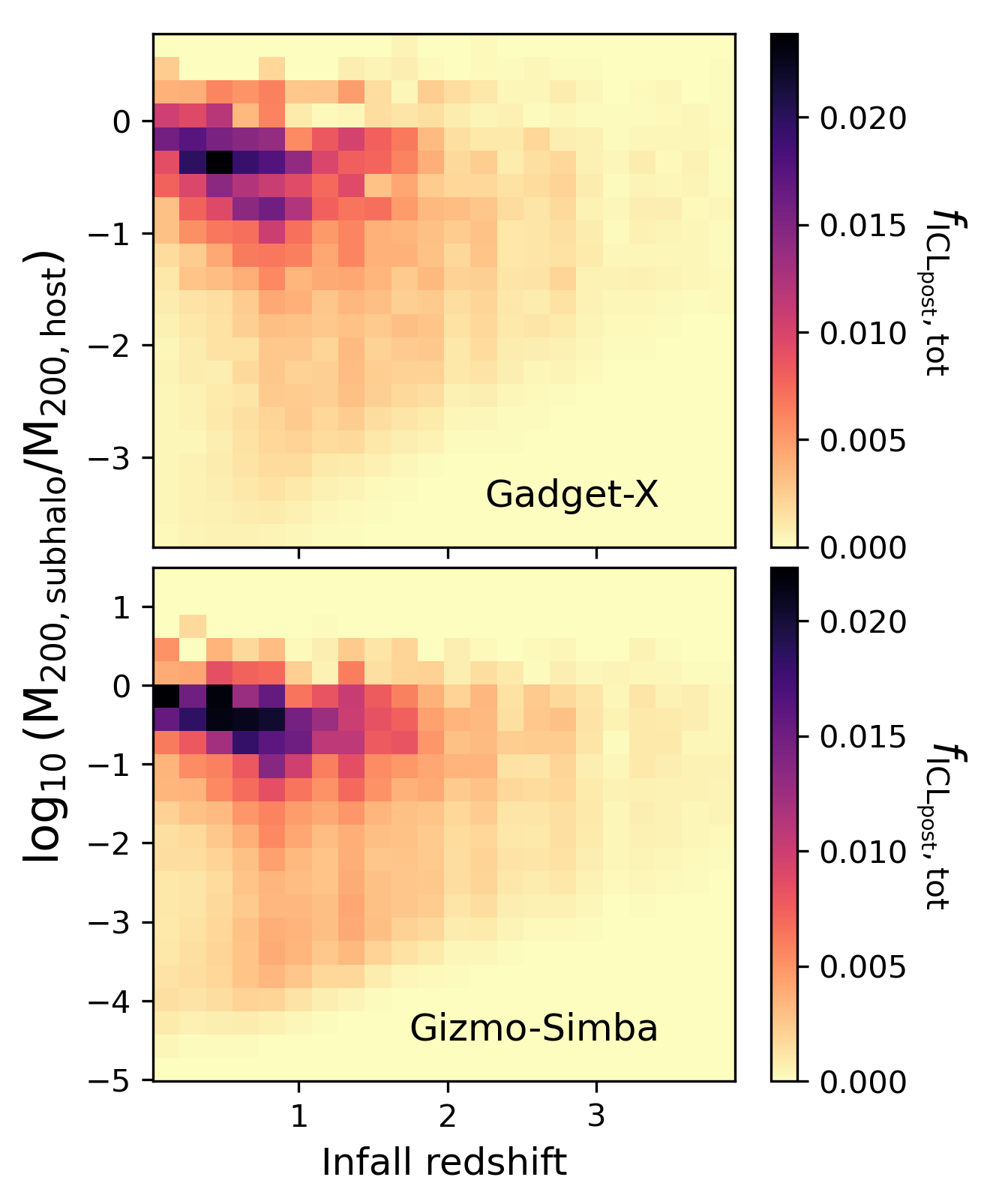}
  \caption{Similarly to \Fig{fig:contrib-2dhist}, 2D histogram showing the distribution of the accreted halos that build the ICL$_\mathrm{post}$, top for \textsc{Gadget-X} and bottom for \textsc{Gizmo-Simba}. Here, the colouring is weighted by the fraction of mass each halo contributes to the total ICL$_\mathrm{post}$ summed across all clusters. The $x$-axis shows the mass ratio of the halo to the host at infall, and the $y$-axis shows the corresponding infall redshift.}
  \label{fig:contrib-final}
\end{figure}

For further investigation into the halos that build the ICL of clusters, we present in \Fig{fig:contrib-2dhist} different density plots with the scatter between some of their properties. Top row is for \textsc{Gadget-X}, bottom one for \textsc{Gizmo-Simba}. In this figure, we are showing 2D histograms coloured by the number of accreted halos within each region of the plot, stacking the 324 clusters together. The scale in the colourbar is the same for the two simulations, highlighting that, as we already saw in \Fig{fig:n-events}, the number of accreted halos is higher for \textsc{Gizmo-Simba}. 
The first column shows the infall redshift of the contributing halos against the mass ratio to the host cluster mass at this time. We can see that there is a positive correlation, such that halos that infall earlier have a mass ratio to the host that is closer to 1. This is quantified by the Spearman correlation coefficient, displayed in the top-left corner of each panel, which confirms the strength and significance of the correlation. This result aligns with expectations from the $\Lambda$CDM hierarchical formation scenario, in which major mergers are more frequent at higher redshifts (see, e.g., Fig. 2 in \citealp{Nuza2024} for the galaxy cluster major merger rate as a function of redshift in \textsc{The Three Hundred} simulations). 
In the second column of \Fig{fig:contrib-2dhist}, we plot in the $y$-axis the mass contribution to the cluster's ICL of each halo, computed as in \Fig{fig:n-events}, $\mathrm{M_{contrib,halo}}/\mathrm{M_{ICL_{post}}}$, as a function of the infall redshift. The trend is weaker now, as quantified by the Spearman correlation coefficient in the bottom corner, but we again see a positive correlation such that earlier infallers, which are less frequent, contribute a higher fraction to the ICL of the cluster. 
In the third column, the $y$-axis still shows the contribution from each halo but as a function of the mass ratio to the host cluster. Consistently with expectations and with the previous scenario, halos with a ratio closer to 1 are those that contribute the most to the ICL of the cluster, although the most common events are those that bring around $10^{-3}$ to $10^{-4}$ of the ICL mass and have a mass ratio of similar value. While the correlation between the mass contributed and the mass ratio is strong ($r_s \sim 0.6$), the figure shows a large scatter. This scatter likely reflects the importance of additional factors such as the orbits of the accreted halos. For instance, halos on more elliptical orbits experience stronger tidal disturbances during close passages, potentially leading to larger contributions to the ICL. We leave a detailed study of halo orbits and stripping processes for a follow-up paper.

\subsection{Contribution to post-stripped ICL}

To sum up the conclusions from \Fig{fig:contrib-2dhist} and further establish the general scenario for the formation of the ICL in \textsc{The Three Hundred} clusters, in \Fig{fig:contrib-final} we show a 2D histogram between the mass ratio of the infalling halo to the host and the infall redshift, with the density weighted by the contribution to the total post-stripped ICL of clusters. This means that, instead of plotting the number of accreted halos as in \Fig{fig:contrib-2dhist}, we now weigh each event by the mass it contributes to its cluster's ICL, $\mathrm{M_{contrib,halo}}/\mathrm{M_{ICL_{post}}}$. We stack the accreted halos for all the clusters and thus divide the weights by the total number of clusters in the sample, 324, leading to the colourbar in the right of \Fig{fig:contrib-final}. 

For both simulations (\textsc{Gadget-X} on top, \textsc{Gizmo-Simba} on the bottom), \Fig{fig:contrib-final} shows that the dominant contribution to the ICL$_\mathrm{post}$ comes from halos that infall at redshift around 1 or below, with mass ratios between 1:10 and 1:1. While \Fig{fig:contrib-2dhist} showed that, individually, early infallers contribute more mass per halo, the more frequent later infallers accumulate to a greater total contribution. Halos accreted earlier, until redshift around $\lesssim 2$, with mass ratios above 1:10, also contribute significantly to the ICL of clusters. The remaining ICL$_\mathrm{post}$ mass arises from numerous minor events, halos with mass ratios below 1:10 and down to $10^{-3}$ or $10^{-4}$, typically infalling between $z \sim 3$ and $z \sim 1$. As in \Fig{fig:relcontrib-cum}, we see a small contribution from mergers with a ratio $>1$. In the third column of \Fig{fig:contrib-2dhist} we have seen that these events are not very common (see density of halos), but they bring a very significant fraction of the ICL, and so in the end they can be fairly relevant for the assembly of the ICL of massive clusters. 

\subsection{Discussion}

In this section we have focused on the `post-stripped' component of the ICL, made of stars that were brought into the cluster within a halo and, after that, became unbound from it and bound to the cluster's potential. We have seen that these contributing halos are generally massive halos. In fact, 65 to 80 per cent of the ICL mass is brought in by infalling objects with stellar mass M$_* > 10^{11}$ M$_\odot$. Looking instead at the total mass of the infalling objects and computing the mass ratio to the cluster's mass, we show that a median of around 35 per cent of the ICL comes from major mergers. As for the redshift distribution, the infall time of these objects peaks at $z \sim 1$, with a tail that reaches until $z \sim 4$. 
Considering each contributing halo an `event' that takes part in the assembly of the ICL, we have computed the ICL fraction contributed by each event. We have seen that, while the most common events are those happening recently and with a very low mass ratio, the ones that contribute the most generally present a high mass ratio and happen at higher redshifts. Adding up all the relative contributions, we conclude that the most relevant contributions to the ICL come from halos with a mass ratio between 1:10 and 1:1, infalling at redshift $\lesssim 1$.

Comparing to previous theoretical works, our results appear consistent with the general trends, while also extending the analysis to a larger sample of more massive clusters. Based on semi-analytic models of galaxy formation and $N$-body simulations, \citet{Contini2014} found that around 68 per cent of the ICL mass comes from satellites more massive than $10^{10.5}$ M$_\odot$. This is in agreement with the prior analytical model by \citet{Purcell2007}, which shows that the ICL is dominated by material liberated from galaxies with M$_* \sim 10^{11}$ M$_\odot$. \citet{Martel2012} find slightly smaller galaxies, with $10^{9}$ M$_\odot$ $ < $ M$_*$ $ < 10^{10.5}$ M$_\odot$ to be the major contributors to the ICL ($\sim 60\%$). All these works agree that the contribution from dwarf galaxies is negligible, even if they dominate in terms of numbers, similarly to our results in \Fig{fig:contrib-2dhist}, where we show that the most common and less contributing events are those with lower mass ratio. Further work with semi-analytic models by \citet{Contini2019}, comparing typical colours of the ICL with those of satellites with different masses, concludes that, at earlier times ($z \sim 1$), the most important contributors to ICL are galaxies with $9 < \log \mathrm{M_*/M_\odot} < 10$, while at lower redshifts ($z \sim 0.5$ and below), more massive galaxies (up to $10^{11}$ M$_\odot$) contribute most. 
This is in agreement with our \Fig{fig:contrib-final}, which shows that objects with a higher mass ratio to the host contribute more at later times. 

Working with hydrodynamical simulations of galaxy clusters, \citet{Brown2024} find that dominant contributors to ICL are galaxies with stellar mass between $10^{10.5}$ and $10^{11}$ M$_\odot$ at their infall time. \citet{Ahvazi2024progs}, also with hydrodynamical simulations but with less massive clusters and massive groups, find 50 per cent of the ICL to come from galaxies more massive than $10^{10}$ M$_\odot$. These works also agree that the dispersion between clusters is wide, which is related to the ICL being formed from very massive objects, so that for an individual cluster the situation can vary greatly. Similarly, \citet{Cooper2015} show that the ICL is very sensitive to contributions of progenitors with a mass ratio of 1:10 (see also \citealp{Harris2017}), matching our \Fig{fig:contrib-final} results, that highlight the contribution from major mergers to the ICL. In contrast to this, analysing mock images from hydrodynamical simulations, \citet{Tang2023} find that the contribution from most massive galaxies to the ICL is minimal, and the dominant ones are within $8 < \log \mathrm{M_*/M_\odot} < 10$. 
With a different set of simulations, \citet{Chun2023} find that half of the ICL and BCG components comes from objects with $\mathrm{M_*} > 10^{11}$ $\hMsun$ in unrelaxed clusters (and from slightly less massive objects for relaxed clusters). In a follow up work, \citet{Chun2024} confirm that typical ICL progenitors fall into the clusters with intermediate mass ($10 < \log \mathrm{M_*/M_\odot} < 11$), with high mass infallers being more important for dynamically unrelaxed and more massive clusters. 

Our results regarding the mass of the contributing objects, as shown in \Fig{fig:relcontrib-cum}, are overall in agreement with previous works (except for \citealp{Tang2023}). Although our objects are in general more massive (we find a median of $10^{11.7}$ M$_\odot$), it is important to note that our host clusters, with M$_{200} > 6.3 \cdot 10^{14}$ M$_\odot$, are also the most massive across all the mentioned studies, and so it is expected that they are formed by more massive contributors \citep[e.g.][]{Chun2024}.

\section{Conclusions} \label{sec:conclusions}

With the advent of new and modern telescopes, such as Rubin, Euclid and JWST, the field of the low surface brightness universe and, specifically, of the ICL, has attracted the interest of the scientific community. New studies show very promising avenues to extract all the information encoded in the ICL, for instance the first ICL study of a cluster with JWST data \citep{MontesTrujillo2022}, the study of the ICL of the Perseus cluster using Euclid Early Release Observations \citep{Kluge2024} or the first spectro-photometric analysis of the entire ICL in a cluster with joint JWST, HST and MUSE data \citep{Jimenez-Teja2024}. In this context, it becomes essential to have theoretical models that can explain the observations and describe the different astrophysical processes involved.
In our previous work \citep{Contreras-Santos2024}, we introduced the ICL of \textsc{The Three Hundred} clusters, showing that it is overall consistent with observations and presenting some of its properties at present day. In this paper, we now provide a theoretical scenario for the formation of the ICL in very massive clusters. In order to do this, we start with the ICL at $z=0$ and trace its particles back in time. Following these particles through the different snapshots of the simulations, we study when, where and how they assemble into the ICL component of the simulated clusters.

\textsc{The Three Hundred} project consists of a suite of 324 spherical regions centred on the most massive clusters found in a prior dark matter-only cosmological simulation. These regions, of 15 $\hMpc$ in radius, have been resimulated including full hydrodynamics in two different implementations: \textsc{Gadget-X}, which uses a modified version of the non-public \textsc{Gadget3} code \citep{Murante2007,Rasia2015}; and \textsc{Gizmo-Simba}, performed with the \textsc{Gizmo} code \citep{Hopkins2015} and the galaxy formation subgrid models from the \textsc{Simba} simulation \citep{Dave2019}. The initial conditions are the same for both resimulations, and so we can consistently compare the resulting clusters, which is interesting to test the robustness of the predictions and understand the effects of the different modelling. Our sample constitutes a mass-complete sample within the range $14.8< \log(M_{200}/M_\odot)<15.6$ at $z=0$. With 324 clusters within this mass, it is a large set of galaxy clusters, similar in statistics to the TNG-Cluster \citep{Nelson2024}, MillenniumTNG \citep{Pakmor2023} or FLAMINGO \citep{Schaye2023} simulations, but much larger than other cluster simulations such as Cluster-EAGLE \citep{Barnes2017} or DIANOGA \citep{Bassini2020}, with one order of magnitude less clusters. 

The main findings of this work can be summarised as follows:

\begin{itemize}
    \item Half of the ICL present day mass was already in the ICL at redshifts between $z \sim 0.2$ and 0.5. At redshift $z \sim 1$, between 10 and 30 per cent of the final ICL was in place (see \Fig{fig:assembly-formation}, top). Both codes show perfect agreement on this, although the stars that were to become ICL in \textsc{Gadget-X} were formed significantly later than those in the \textsc{Gizmo-Simba} run. Besides, both simulations show that clusters that formed earlier also assemble their ICL earlier (\Fig{fig:assembly-formation}, bottom), highlighting the close connection between the ICL component and the whole cluster (including the dark matter halo).
    
    \item We define in-situ stars as those formed directly in a progenitor of the central halo and not in any other object. \textsc{Gadget-X} shows a fraction of in-situ stars between 5 and 20 per cent of the total ICL, while \textsc{Gizmo-Simba} shows a median value of 1.5 per cent, with the highest values being around 5 per cent.
    
    \item Studying the distance from the in-situ stars to the cluster centre at the time of their formation (\Fig{fig:insitu-distance}), we find that the in-situ stars in \textsc{Gizmo-Simba} can be associated to the BCG, being formed in the centre and ejected later into the ICL. In \textsc{Gadget-X}, around 35 per cent of the in-situ stars are formed at a distance $r \gtrsim 0.2 R_{200}$ from the cluster centre. The origin of these particles in \textsc{Gadget-X}, and whether it is physical or numerical, remains unclear, since an in-depth investigation is out of the scope of this paper, but we find them to be associated with gas-dominated blobs likely formed due to the disruption or ram-pressure stripping of recent infalling halos.
    
    \item We classify the remaining ICL stellar particles (ex-situ stars) based on the halo they belonged to before being accreted to the ICL and on when this happened. The major contribution is from `post-stripped' stars ($\sim 90\%$ of ex-situ stars, see \Fig{fig:stars-classification}), that fell into the cluster inside another halo, and later became unbound from it and bound to the cluster halo. Other minor contributions are from stars formed in a subhalo already inside the cluster and then stripped from it, stars accreted into the ICL from a halo that was still outside the cluster halo before its infall and stars contributed by halos that never fall into the cluster.
    
    \item Focusing only on the post-stripped ICL, we find that, depending on the cluster, 65 to 80 per cent of the total ICL is brought in by objects with M$_* > 10^{11}$ M$_\odot$ at their infall time. In terms of total mass ratio of these objects to the cluster, major mergers involve, in median, around 35 per cent of the mass in the post-stripped ICL, ranging from around 15 to 55 per cent for the different clusters. 

    \item Studying these events that make up the post-stripped ICL in more detail, we show in \Fig{fig:contrib-2dhist} that the most common events are recent infallers with a very low mass ratio (very minor mergers or slow accretion). However, the higher the infall redshift and the mass ratio of the halo to the host cluster, the higher is the fraction that this halo contributes to the ICL. Stacking all the clusters together, in \Fig{fig:contrib-final} we see that the main channel to build ICL is from minor or major mergers at $z \leq 1$, in agreement with the previous results. Halos with a mass ratio close to 1, but with higher infall redshift, between 1 and 2, are also contributing a relevant fraction, while the remaining mass comes from very minor mergers or accretion at redshifts $1 \lesssim  z \lesssim 3$.
    
    \item Finally, in \Fig{fig:n-events} we present the minimum number of accreted halos needed to assemble any fraction of the post-stripped ICL. For both implementations of the hydrodynamics, a minimum of 2 to 8 halos is needed to build 50 per cent of the ICL. For the whole ICL the results differ notably, with \textsc{Gadget-X} requiring between 75 and 150 halos, and \textsc{Gizmo-Simba} 300 to 600. The mass of these halos also differs from one simulation to the other, with \textsc{Gizmo-Simba} showing a more important contribution from smaller subhalos that infall into the main cluster (\Fig{fig:relcontrib-cum}). 
\end{itemize}

In summary, we find that the build-up of the ICL is mainly due to very massive galaxies as well as major mergers with other massive groups or clusters, preferentially happening at redshifts below 1. We note that we have not distinguished between mergers and stripping throughout this work, that is, we are not taking into account if the contributing halos end up being disrupted or they survive as subhalos of the host cluster. We also do not check if the halos merge first with the BCG and then the stars are moved to the ICL. These aspects will be the focus of a forthcoming follow-up study, where we will analyse in more detail the properties of contributing halos. In particular, we will investigate whether the contributing stars were already part of a diffuse component in their original halos, as well as characterise the stripping process itself—examining orbital parameters, cosmic web environment, stripping timescales, the fraction of mass lost, and related dynamical properties.

\begin{acknowledgements}
    This work has been made possible by \textsc{The Three Hundred} (\url{https://the300-project.org}) collaboration. The simulations used in this paper have been performed in the MareNostrum Supercomputer at the Barcelona Supercomputing Center, thanks to CPU time granted by the Red Espa\~{n}ola de Supercomputaci\'on. As part of \textsc{The Three Hundred} project, this work has received financial support from the European Union’s Horizon 2020 Research and Innovation programme under the Marie Sklodowskaw-Curie grant agreement number 734374, the LACEGAL project. 
    ACS, AK, GY, IAA and CDV thank the Ministerio de Ciencia e Innovaci\'{o}n (MICINN) for financial support under research grants PID2021-122603NB-C21 and PID2021-122603NB-C22. AK further thanks My Bloody Valentine for loveless. WC is supported by the Atracci\'{o}n de Talento Contract no. 2020-T1/TIC19882 granted by the Comunidad de Madrid in Spain, and the science research grants from the China Manned Space Project. He also thanks the Ministerio de Ciencia e Innovaci\'{o}n (Spain) for financial support under Project grant PID2021-122603NB-C21 and HORIZON EUROPE Marie Sklodowska-Curie Actions for supporting the LACEGAL-III project with grant number 101086388. SEN is a member of the Carrera del Investigador Cient\'{\i}fico of CONICET. He acknowledges support from CONICET (PIBAA R73734) and Agencia Nacional de Promoci\'{o}n Cient\'{i}fica y Tecnol\'{o}gica (PICT 2021-GRF-TI-00290).

\end{acknowledgements}


\bibliographystyle{aa}
\bibliography{archive}




\label{lastpage}
\end{document}